\documentclass[iop, apj]{emulateapj}
\usepackage{apjfonts}
\usepackage{url}

\def\etal{et al.}

\def\kms{km~s$^{-1}$ }

\def\nh{$N_{\rm H I}$}

\def\muG{$\mu$G}

\def\cm2{${\rm cm}^{-2}$}
\def\cm3{${\rm cm}^{-3}$}

\def\ciis{C II$^{*}$}

  \slugcomment{Accepted by the Astrophysical Journal}

\shorttitle{Polarization of 21cm Absorption Line in 3C 286}
\shortauthors{Wolfe, Jorgenson, Robishaw, Heiles, Prochaska}

\begin{document}

\title{SPECTRAL POLARIZATION  of the
REDSHIFTED 21 cm ABSORPTION LINE TOWARD 3C 286}

\author{Arthur M. Wolfe,\altaffilmark{1} Regina A. Jorgenson,\altaffilmark{2} 
Timothy Robishaw, \altaffilmark{3} Carl Heiles, \altaffilmark{4} J. Xavier Prochaska, \altaffilmark{5}}
\email{awolfe@ucsd.edu} 

\altaffiltext{1}{Department of Physics and Center for Astrophysics and Space Sciences, UCSD, La Jolla, CA 92093, USA}
\altaffiltext{2}{Institute of Astronomy, University of Cambridge, Madingley Road, Cambridge, CB3 OHA, UK}
\altaffiltext{3}{Sydney Institute for Astronomy, The University of Sydney, NSW 2006, Australia}
\altaffiltext{4}{Department of Astronomy, University of California, Berkeley, CA 95064}
\altaffiltext{5}{Department of Astronomy \& Astrophysics, UCO/Lick Observatory, 1156 High Street, University of California, Santa Cruz; CA 95064}

\begin{abstract} 

A re-analysis of the Stokes-parameter spectra obtained
of the $z$=0.692 21 cm absorption line toward
3C 286 shows that our original claimed detection
of Zeeman splitting by a line-of-sight
magnetic field,
$B_{\rm los}$ = 87 {$\mu$}G is incorrect. Because of
an insidious software error, what we reported as
Stokes $V$  is actually Stokes $U$: the revised
Stokes $V$ spectrum indicates a 3-$\sigma$
upper limit of  
$B_{\rm los} \ <$ 17 {$\mu$}G. 
The correct
analysis reveals an absorption feature in 
fractional polarization 
that is offset in velocity 
from the Stokes $I$ spectrum 
by $-$1.9 km s$^{-1}$. The polarization position-angle
spectrum shows a dip that is
also significantly offset from the Stokes $I$ 
feature, but at a  velocity
that differs slightly from the absorption
feature in fractional polarization. 
We model the absorption feature with 
3 velocity components against the core-jet
structure of 3C 286. Our $\chi^{2}$ minimization
fitting results in components with differing
(1) ratios of H I column density to spin temperature,
(2) velocity centroids, and (3) velocity dispersions.
The change in polarization position 
angle with frequency implies incomplete
coverage of the background jet source by the
absorber. It also implies a spatial
variation of
the polarization position angle across the jet
source, which is observed at frequencies higher
than the 839.4 MHz absorption frequency. 
The multi-component structure of the
gas is best understood in terms of components with
spatial scales of $\sim$ 100 pc comprised of
hundreds of
low-temperature ($T$ $\le$ 200 K)  clouds
with linear dimensions of $<<$  100 pc. 
We conclude
that previous attempts to model the foreground
gas with a single uniform cloud are incorrect.

\end{abstract}


\newpage

\section{INTRODUCTION}

In order to detect magnetic fields in galaxies with significant
redshifts,
we began a search for Zeeman splitting of 21 cm
absorption lines arising in damped lyman alpha systems (DLAs) toward
background quasars (Wolfe {\etal} 2005) 
selected to be radio bright.  The detection of Zeeman
splitting yields the strength, direction, and redshift of the magnetic
field, which is an advantage over measurements of Faraday
rotation, 
which yield rotation measures
accompanied with huge uncertainties in the strength and redshift of the
$B$ fields along the sightlines to background quasars (Kronberg {\etal} 2008).
In Wolfe {\etal}
(2008) we reported the detection of Zeeman splitting in the $z$=0.692
DLA toward 3C 286. The purpose of this paper is to show that the
reported detection was erroneous. In \S \ref{obs2} we describe the
errors that led to the claimed detection and describe our new data
reduction procedures. In \S \ref{obs3} we correct
the previous errors and give a correct description of the polarization
spectra. The new spectra  are shown in \S \ref{proper}.

In the absence of detectable Zeeman splitting we can
only set an upper limit to the magnetic field in the
$z$=0.692 DLA toward 3C 286. However, in this paper
we discuss how the
spatial distribution of the strong linear polarization 
of 3C 286 itself can be used in combination with 
observed spectral variation of linear
polarization across the absorption profile of
the redshifted 21 cm line to infer physical properties
of
the absorbing H I gas. We
demonstrate how comparison between   
properties of the  linearly polarized and unpolarized
absorption spectra  provides new information about the spatial
structure of this gas on scales of $\sim$ 100 pc.
We use the new data to determine both
the kinetic temperature and hyperfine spin temperature
of the gas, which in turn tells us about its thermal state.
We use these results to evaluate the standard assumption
of a large-scale, uniform cloud. 
We start by constructing models to describe
the  re-analyzed spectra in terms of multiple
clouds distributed across the spatially extended
structure of 3C 286.
In \S \ref{cloudstructure}.1 we use
VLBI maps to exhibit the core-jet structure of 3C 286 
near the 21 cm absorption frequency. In \S \ref{cloudstructure}.2
we describe a two-cloud configuration to model the  absorption feature.
We use a chi square minimization technique
to fit the model to the Stokes $I$ spectrum
and the fractional polarization spectrum and show that the model
does not work. In \S \ref{cloudstructure}.3 we show how
an adjustment in covering factors and the presence of velocity
gradients in the cloud toward the polarized jet source results
in a three-component model that provides an adequate fit
to the Stokes $I$, fractional polarization, and polarization
position-angle spectra. 
The implications
of the results are discussed in \S \ref{discussion}.
All the results are then summarized in \S \ref{summary}. 
 
Throughout this paper we use a WMAP cosmology
in which ($h, {\Omega_{\rm m}}, {\Omega_{\Lambda}}$)
=(0.7, 0.3, 0.7).

\section{OBSERVATIONS} \label{obs2}

We used the Green Bank Telescope in 2007 to observe the
$z=0.692$ 21 cm  absorption spectrum against 3C 286. We observed all four
Stokes parameters simultaneously using the digital FX Spectral
Processor, which provides all the necessary self- and cross-products;
here ``FX'' means that first it Fourier transforms the input signal and
then multiplies the voltage spectra with appropriate phase shifts. This
technique is described in detail by Heiles et al.\ (2001) and Heiles
(2001).

Our original paper reported a very statistically significant
detection of circular polarization (Stokes $V$) for the 21-cm line DLA
system against 3C 286. As Zeeman splitting, it translates to a
line-of-sight field strength of 84 $\pm$9 $\mu$G. Subsequently, 
after re-observing 3C 286 in January and March 2009, we found
that this detection is incorrect, basically because what we reported as
Stokes $V$---circular polarization---is really Stokes $U$---linear
polarization. The discussion below explains the origin of this error,
which is a matter of a misplaced $90^{\circ}$ of phase. 

\subsection{Calibrating a Local Noise Source as a Secondary Calibrator 
Standard} 
 
The sky electric field is sampled by probes in the telescope feed. In our
observations with the GBT, these two native polarizations are very close
to being orthogonal linears. For convenience we denote the sampled
voltage spectra (i.e., the Fourier transforms of the sampled voltages)
as $X$ and $Y$. Apart from small corrections for nonorthogonality and
other coupling, the Stokes parameters are obtained as follows:
 
\begin{mathletters}
\begin{equation}
I = XX + YY
\end{equation}
\begin{equation}
Q = XX - YY
\end{equation}
\begin{equation}
U = 2 {\cal R} (X Y) 
\end{equation}
\begin{equation}
V = 2 {\cal I} (X Y) 
\end{equation}
\end{mathletters}

\noindent We tacitly assume that the voltage spectral products are
time averages. Here $\cal R$ and $\cal I$ mean the Real and Imaginary
parts, respectively. To clarify these equations and their discussion, we
neglect the small corrections for nonorthogonality; these corrections
are embodied in the feed's Mueller matrix and are determined as
discussed in \S \ref{muellercorr}.

%There are many electronic components between the feed and the Spectral
%Processor. These increase the signal amplitude by a gain factor $G_e$;
%this is necessary because the voltage inputs to the Spectral Processor
%must be billions of times larger than those sampled by the feed. These
%components also add phase to the signal; 

In a heterodyne radio receiver the signal from the sky is amplified by 
mixing (multiplying) the sky signal by one or more locally-generated
signals of known amplitude and phase. The resulting signal product,
the gain $G_{e}$, is a complex number in which
the amplitude is the real part  and the phase the imaginary part.
Between the receiver and the Spectral Processor there are
a variety of electromagnetic components that alter the phase 
of the signal. Specifically, there are cables and optical fiber,
each of which adds phase proportional to $2 \pi L
\over \lambda$, where $L$ is the length and $\lambda$ the
wavelength of the signal in the particular cable. Thus the gain $G_e$ is
a complex number. Moreover, 
some of these electromagnetic components have their own polarization
properties that
differ for $X$ and $Y$, and as a result we have $G_{e,X}$ and
$G_{e,Y}$. For the complex portion of these gains, the only part of
concern is the phase angle difference

\begin{mathletters}
\begin{equation} \label{phieqn}
\phi_e = \phi_{e,Y} - \phi_{e,X}
\end{equation}

\noindent where

\begin{equation} 
\tan \phi_{e,Y}  = { {\cal I}(G_{e,Y}) \over {\cal R}(G_{e,Y}) }
\end{equation}
\end{mathletters}

\noindent and a similar equation for $X$. 

These complex gains must be calibrated and their effects removed. The
standard technique is to inject perfectly correlated noise at the feed
from a noise source, commonly called the ``cal''. Like the sky signal,
the correlated noise source injects voltages into the feed;
%3, which we denote by $X_c$ and $Y_c$; 
unlike the sky signal, the cal's noise is
perfectly correlated and corresponds to 100\% polarization.  We compare
the cal's signal with that of a standard polarized calibration source of
known flux to determine the equivalent antenna temperatures of the
cal. We also compare the relative phase of the cal with that of the
source, which relates the polarization of the cal to that of the sky. 

In essence, we determine the gain of the cal in $X$ and $Y$ relative to
the sky signal; these gains are complex and change
with frequency. We assume the
cal's properties to remain constant with time. In subsequent
observations, we use the cal's deflection to determine the complex
electronic gains $G_{e,X}$ and $G_{e,Y}$.

\subsection{Using the Cal to Calibrate the Complex Gains}

Consider the frequency dependence of equation \ref{phieqn}. The
individual phase terms on the right-hand-side come from electronic
components and from cables. The phase delays of most electronic
components change relatively slowly with frequency. However, for cables,
the phase delay is $2 \pi L \over \lambda$ and the phase difference in
equation \ref{phieqn} is $\Delta \phi = {2 \pi \Delta L \over \lambda}$,
where $\Delta L$ is the length difference between $Y$ and $X$. In our
experience, $\Delta \phi$ varies fairly rapidly with frequency.  In particular,
using the techniques described by Heiles (2001), for the current
observations we find

\begin{equation}
{d \Delta \phi \over d{\nu}} \approx { 20 \ \ {\rm deg \ MHz^{-1}} }
\end{equation}

We have well-tested software that reliably performs this determination,
and its correction (e.g.\ Heiles \& Troland 2004). Unfortunately,
however, this software had a bug, which we discovered on June 5,
2009 while we were analyzing both the 2007 and 2009 data. 
If we fed the software an array of spectra to correct, it
performed correctly. But if we fed the software just a single spectrum,
it didn't apply any correction at all. Before the current observations,
we had never used the software on individual spectra.

During our original 2007 observations this phase delay happened to
average about 91 degrees at band center with a slope of about 20
deg/MHz; the phase delay fluctuated by at most a few degrees for all of
our spectra. 
For
our bandwidth of 625 kHz, the slope doesn't matter much. Thus, the phase
difference of $\sim 90$ degrees effectively interchanges the real and
imaginary parts of the cross-correlation product. In turn, this changes
the derived Stokes $V$ into Stokes $U$, and vice-versa.
Fortunately, in our 2009 observations the 
phase delay  was about 30 degrees, which
alerted us to the error in the 2007 data.  
Although the precise cause of the change in phase
delay between 2007 and 2009 is unknown, we note
that a change in cable length in either
polarization over the entire signal path
between the feed and correlator could result in
such a change.

\subsection{The Mueller Matrix} \label{muellercorr}
We corrected for polarization impurities in the system by using the
Mueller matrix formulation given by Heiles et al.\ (2001). The Mueller
matrix relates the measured auto- and cross-correlation products to the
actual Stokes parameters. With this
procedure, one observes a linearly polarized calibrator over a range of
parallactic angle $PA$ to derive seven parameters that describe the
complex gains and coupling coefficients of the receiver system
components. One then derives the Mueller matrix coefficients from
algebraic combinations of these coefficients.

In the present paper, we observed 3C 286. This source is, for most
purposes at cm wavelengths, the ``gold standard'' polarization
calibrator. Specifically, the fractional polarization
and polarization position angle of 3C 286 have been stable
over four decades  of observations across
a wide range of frequencies above $\sim$ 1400 MHz (Tabara \& Inouye
1980; B. Gaensler 2009 priv. comm.). While these properties
have not been established at lower frequencies,
the dominance of the large-scale jet at these  frequencies
argue against any time variations of these quantities
at 840 MHz. 
Hence we had the fortunate circumstance of observing a
source that also happens to be an excellent polarization standard. For
about half of our observing days, we had enough $PA$ coverage to derive
the Mueller matrix from our observed data for that particular day; thus,
we could use our observed astronomical data to also determine the
Mueller matrix elements---a form of ``self calibration''. The derived
corrections were closely the same from one day to another, although there
were small differences. For any particular day's observations, we used
the matrix that was closest in time to that day; for about half the
days, this matrix was obtained on that very day. This gives us great
confidence in our polarization calibration.

Normally, one determines a single Mueller matrix for the observed band
by averaging over the whole band. This leads to small calibration errors
that change with frequency across the band, which in turn leads to small
``baseline curvature'' in the three polarization spectra. The latter are
the fractional polarization $p(v)$, polarization position angle
$\chi(v)$, and Stokes $V$ parameter $V(v)$; 
one normally fits a smooth baseline to these
spectra and applies them as an {\it ad-hoc} correction.

Here, we took a different approach. 3C 286 is such a strongly polarized
source that, for any one day, we could easily determine the seven
receiver system parameters on a channel-by-channel basis. We then fit
their frequency dependencies by performing a
minimum-absolute-residual-sum (MARS) 3rd-degree polynomial
fit\footnote{The MARS fit has the advantage of not responding to the
  spectral features caused by the absorption line.  To our knowledge,
  this technique was invented by P.B.\ Stetson; see his website 
  {\tiny \url{http://nedwww.ipac.caltech.edu/level5/Stetson/Stetson4.html}}.
  Heiles has additional information on this technique; see 
  {\tt lsfit\_2008.ps} on his website 
    {\tiny \url{http://astro.berkeley.edu/\~heiles/handouts/handouts\_num.html}}.
  }. 
   We used these fit
coefficients to derive the Mueller matrix independently for each
channel. With this, the polarization calibration is very accurate for
each channel independently, so that any features in the polarized
spectra are real, characterizing the sky instead of the system. In
particular, it is neither appropriate nor necessary to subtract off an
{\it ad-hoc} baseline in any of the polarized spectra. It is these
accurately-calibrated spectra that we use and plot in this paper. Even
with this self-calibration of each channel independently, there remain
small residual inaccuracies, which result from applying Mueller matrices
obtained on one day to data obtained on another day; this is probably
responsible for the small spectral effects in the Stokes $V$ spectrum in
Figure \ref{newStokes}.

\section{REANALYSIS} \label{obs3}

First, we fixed the above-mentioned software bug. Then we reanalyzed the
data.  We used the same software as before and determined the Mueller
matrices in the usual way (see Heiles et al.\ 2001), as we did before. We
applied the Mueller matrix corrections and the phase difference
correction of equation \ref{phieqn} as we did before. We did three
things differently: \begin{enumerate}

\item We derived the linear polarization by fitting the Parallactic
  angle dependence of Stokes $U$. Stokes $U$ is a cross product and
  produces more reliable results, free of zero offsets, than Stokes $Q$,
  which is the difference between two large numbers. (As it happens, the
  results from Stokes $Q$ are indistinguishable from those of Stokes
  $U$).

\item We reduced the off-line (i.e., the 3C 286 continuum) spectral
  channels to ensure that we obtain the correct results. That is, we did
  not ``subtract off the continuum baseline''. 3C 286 is the
  premier polarization calibrator in radio astronomy, and this ensures
  that our spectral line and continuum results correspond to what is
  known about 3C 286.

The one wrinkle in this is that we have been unable to find a primary
measurement of the polarization properties of 3C 286 at frequencies as
low as ours, 840 MHz. While 3C 286 exhibits essentially zero change with
frequency of polarization position angle above 1.4 GHz, we cannot be
sure that the angle at 840 MHz isn't somewhat different. We did not wish
to address this question and therefore our reported polarization
position angles have an arbitrary zero point.

\item We edited the data much more carefully than before, which
  produced only minor improvements.

\end{enumerate}

\section{THE PROPER DATA: STOKES PARAMETERS} \label{proper}

\begin{figure*}[p!]  \begin{center}
\includegraphics[width=6.0in]{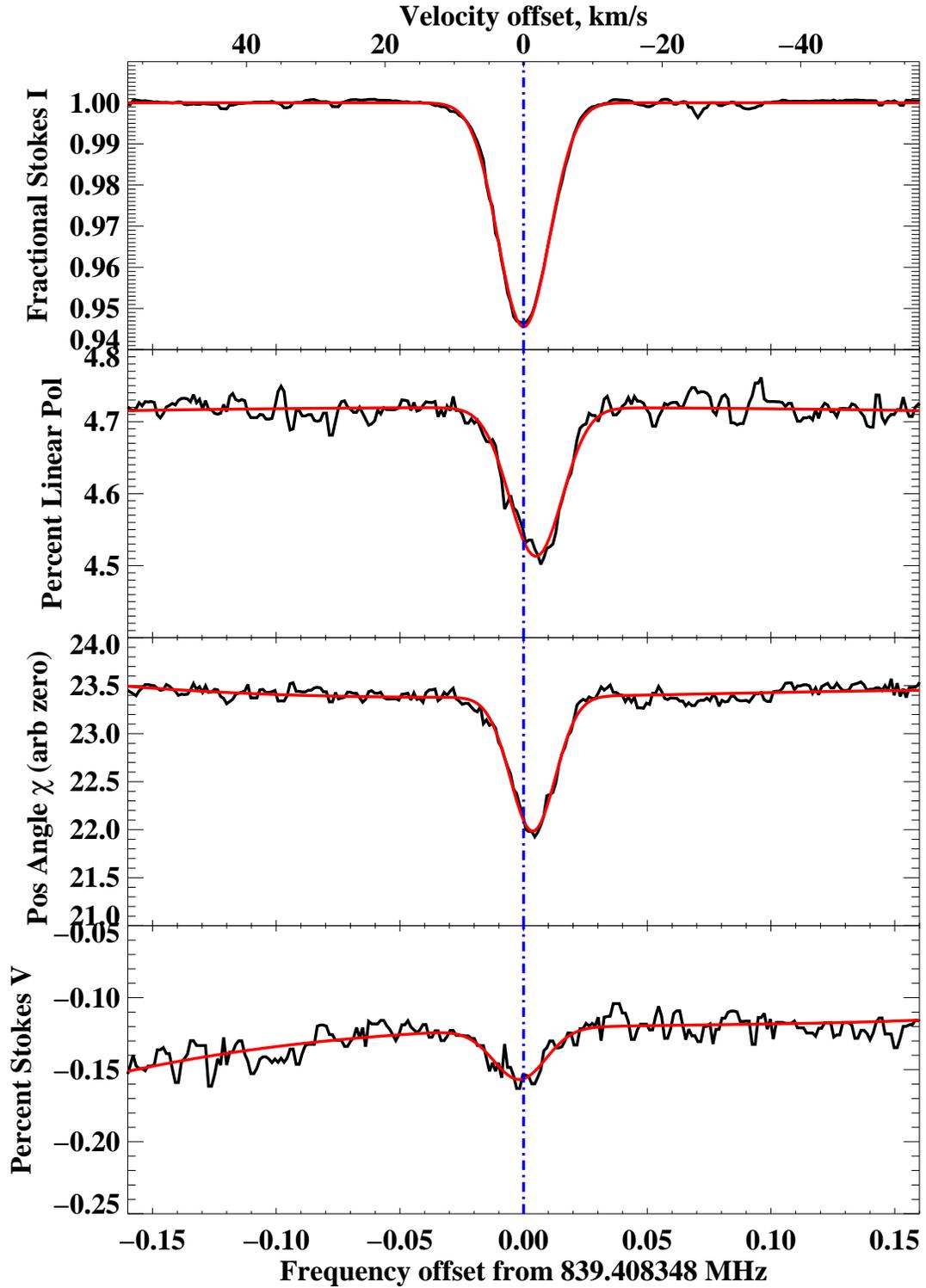} \end{center} \caption{
    Plots of Stokes parameters for 3C 286. From top to bottom, the panels
    are: the fractional absorption spectrum $I(v)/I$; fractional linear
    polarization $p(v)$; position angle of linear polarization $\chi(v)$; and
    fractional circular polarization $V(v)/I$. Fractional spectra are in
    percent and frequencies are barycentric in MHz offset from 839.408348
    MHz. Black curves are data and orange curves are fits to the data.}
\label{newStokes}
\end{figure*}

 Figure \ref{newStokes} shows a four-panel plot of the 
data (black curves),
now reduced properly. 
No ``baseline corrections'' have been subtracted
or applied. The data consists of 12.6 hrs of on-source
integration obtained in 2007. The orange curves
are least squares fits to the data. 
The top panel shows Stokes $I(v)/I$ (where
$I$ is the off-line continuum intensity, which is
independent of velocity $v$), which has an absorption
line with a fractional absorption of about 5\%.  
The frequency centroid determined  by the fit to the Stokes
I spectrum is 839.408348 $\pm$ 0.000046 MHz and
its location is depicted by the vertical dot-dashed
line. The corresponding redshift
$z$=0.69215109$\pm$0.00000009.
The second panel shows
the percent polarization, $p(v)={\sqrt{(U(v)^{2}+Q(v)^{2})}}/I$, and the third panel shows the position
angle $\chi(v)$
(zero point is arbitrary).

The linearly polarized $p(v)$ line also appears in absorption (i.e.\ the
percent polarization goes down in the line) and its fractional polarized
absorption is close to that for Stokes $I$. Its {\it line center}
differs from the Stokes $I$ line center by $5.0 \pm 0.25$ kHz. This $20
\sigma$ difference has high statistical significance.  Taken together,
these two results are consistent with the absorbing gas covering both
nonpolarized and polarized portions of the continuum source image, and
the velocity of the part that covers the polarized portion differing
from that covering the unpolarized portion by 5.0 kHz (which is about
1.8 km/s)\footnote{This shift is what looked like Zeeman splitting
  in our earlier result, where we thought that Stokes $U$ was Stokes
  $V$. If the frequency shift for polarized intensity were reduced with
  the same software as we use to derive Zeeman splitting from Stokes
  $V$, then we would derive a field strength of 280 $\mu$G (instead of
  the original one in the paper of ``only'' 84 $\mu$G). This is because,
  when averaging linearly polarized Stokes parameters over hour angle,
  we properly rotate by the parallactic angle $PA$; in contrast, when we
  averaged our old Stokes $V$ we didn't account for 
  $PA$, so the amplitude was smeared down by improper weighting in
  $PA$).}.

The third panel shows that the line changes the position
angle by about 1.5 degrees at line center, which is consistent with the
absorbing gas covering only a fraction of the polarized source image and
requires that the position angle of polarization of the continuum
changes across the source image.

The fourth panel shows the percentage circular polarization
$V(v)/I$.  The absorption line
exhibits a dip in fractional Stokes $V$
of $0.03\%$.  However, we believe this
apparent circular polarization is meaningless.  The  line is $5\%$
linearly polarized, and some of the total intensity and also the linear
polarization will leak into Stokes $V$.  The observed Stokes $V$
intensity of the line is only about $1\%$ of its linearly-polarized
intensity.  Errors in the Stokes parameters at these levels are not
unexpected, because of the small day-to-day changes in the Mueller
matrix elements.

\section{CLOUD STRUCTURE PRODUCING THE ABSORPTION FEATURE}
\label{cloudstructure}

We now show that
the GBT polarization spectra provide new and unique information
about  the kinematic and spatial
structure of the  gas that gives rise to 21 cm absorption toward
3C 286.  In particular we note the following model-independent
facts: 
(1) The difference between the velocity centroids of
the Stokes $I$ and fractional polarization spectra 
is inconsistent with a single-cloud origin for the absorption.
(2) The Gaussian shape of the absorption profile is plausibly
explained through the cental limit
theorem by the superposition of many small clouds. (3) The
position-angle spectrum reinforces the conclusions drawn
from the fractional polarization, but requires detailed modeling
as described below. Besides
studying the spatial and kinematic structure of the absorbing
gas, we also investigate  
its two-phase structure;
i.e., whether the gas is a cold
neutral medium (CNM) with temperature $T$ $\sim$ 100 K, a
warm neutral medium (WNM) with $T~\sim$ 8000 K (Wolfire {\etal} 1995),
a thermally unstable phase with temperature between these two
extremes, or some combination of any of these. Furthermore,
we  consider clues  to the identity of
the galaxy hosting the absorbing gas.

\subsection{Core-jet structure of the radio source}

\begin{figure}[h!]  \begin{center}
\includegraphics[width=3.0in]{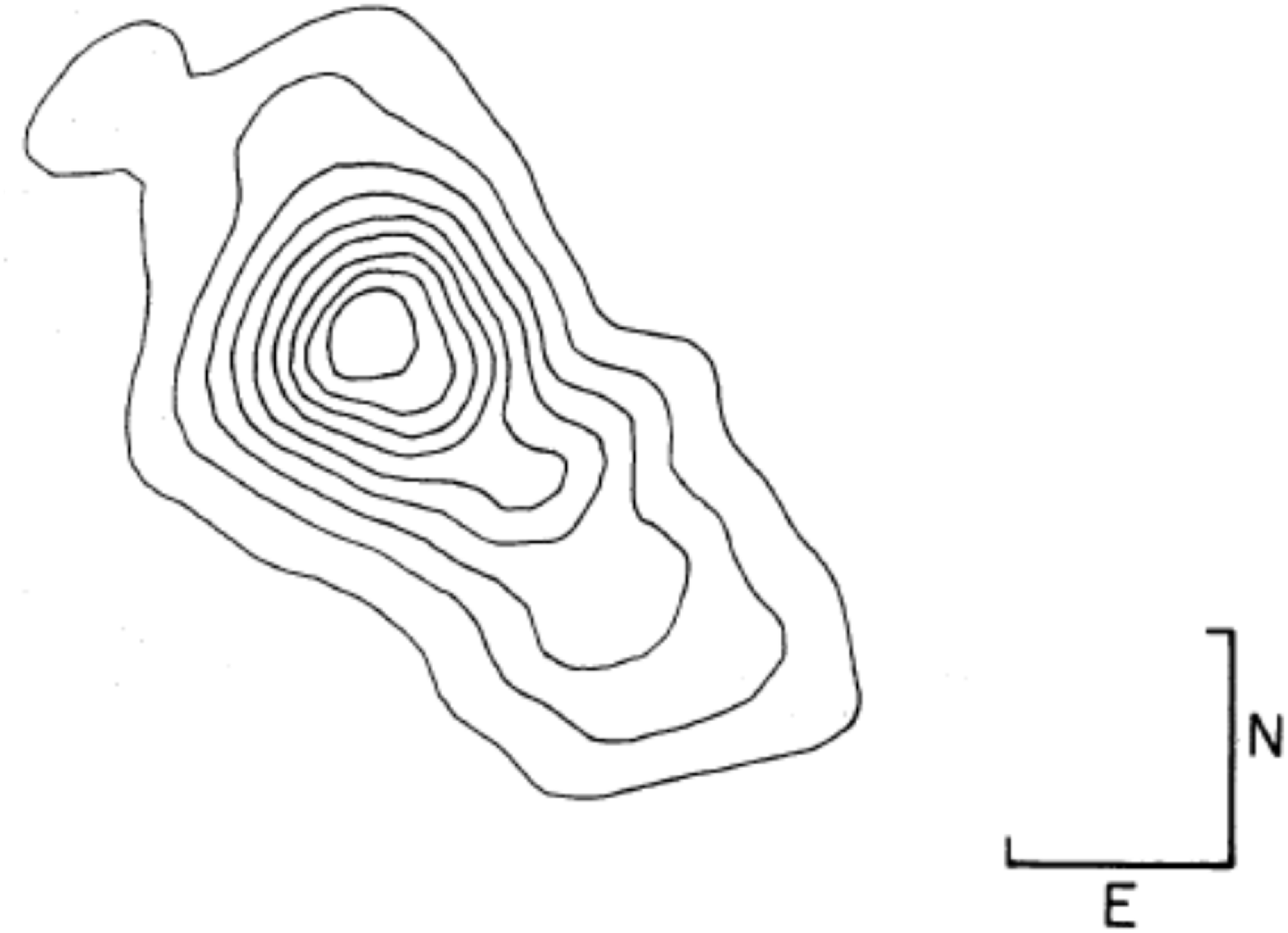}
\end{center} 
\caption{VLBI
map of 3C 286 at 609 MHz (Wilkinson {\etal} 1979). 
Continuum structure exhibits
an unresolved core source to the north east and an extended
jet source to the south west. Orthogonal straight lines are 0.02 arcsec
in length corresponding to 142 pc at the absorber.} 
\label{vlbimap}
\end{figure}

The 21 cm absorption likely forms against a source with the
core-jet configuration shown in Fig.~{\ref{vlbimap}} (Wilkinson
{\etal} 1979). 
This VLBI map,
obtained at 609 MHz, indicates that 3C 286 is an asymmetric source  consisting
of a compact core and an extended jet. The core,
which covers a solid angle, $\Omega_{\rm core} {\approx}10{\times}5$
milli-arcsec$^{2}$ (mas$^{2}$) with major-axis 
position angle equals 42$^{\circ}$, contributes
a flux density $S_{\nu}(609)$=3.5 Jy, while in the case
of the jet $\Omega_{\rm jet}$
=40$\times$15 mas$^{2}$, P.A.=42$^{\circ}$, and $S_{\nu}(609)$=14 Jy.
Since a similar core-jet structure is detected at 329 MHz,
1667 MHz (Simon
{\etal} 1980), 5 GHz (Jiang {\etal} 1996; Cotton {\etal}
1997), and 15 GHz (Kellermann {\etal} 1997, priv. comm.), it
is safe to assume that such a structure is present at 
the absorption frequency of 839.4 MHz. Adopting the
spectral indices measured by Simon {\etal} (1980) of
${\alpha_{\rm  core}}=-0.29{\pm}$0.15 
and $\alpha_{\rm jet}=-0.55$,
we find that $S_{\nu}^{\rm core}$${\approx}$3.2 Jy and
$S_{\nu}^{\rm jet}${$\approx$}11.7 Jy at $\nu$ $\approx$ 840 MHz.

While the polarization structure of 3C 286 has not
been detected at low frequencies, VLBI  measurements
of all four Stokes parameters have been obtained
at 5 GHz (Jiang {\etal} 1996; Cotton {\etal} 1997; Cotton 2010 [priv. comm.]).
These data show that unlike most quasar VLBI sources,
the fractional polarizations, i.e., fraction of
total surface brightness 
in $\sqrt{U^{2}+Q^{2}}$, where $U$ and $Q$
are the Stokes parameters,
of the core 
and jet are comparable. More specifically, the ratio
of jet to core polarized surface brightnesses is about 0.4.

\noindent Nevertheless the ratio of polarized flux densities
$P^{jet}/P^{core}$ $\approx$ 5 owing to
the much larger solid angle subtended by the jet,
where
the polarized flux density $P=\sqrt{U^{2}+Q^{2}}$.
Because of the steeper spectral index of the jet, it is reasonable
to assume that this ratio increases with decreasing
frequency. As a result, we shall ignore 
the core source when computing absorption in Stokes
$U$ and $Q$ parameters, or equivalently in the spectrum
of polarized flux density $P$
(note, $P=pI$) and polarization
position angle $\chi=0.5{\times}$arctan($U/Q$).
We check the self-consistency of this assumption
in $\S$ 7.

\subsection{Two Cloud Model}

We now describe the procedures used to model the Stokes 
parameter spectra.

\subsubsection{Model Description}

\begin{figure}[b!]  \begin{center}
\includegraphics[width=3.0in]{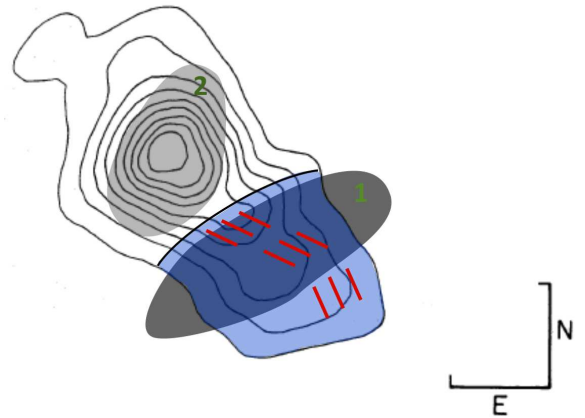} 
\end{center} 
\caption{Foreground clouds (grey) superposed on VLBI
map from Fig.~{\ref{vlbimap}}. Cloud 2 at
NE completely covers core source, while
cloud 1 toward SW partially covers resolved jet source. 
Cloud 1 is shaded darker  to illustrate its
larger 21 cm optical depth.  
Light blue shading  indicates that region
emitting polarized radiation is the jet.  
Thick red bars depict direction of 
E vector polarization. Orthogonal straight lines are 0.02 arcsec
in length corresponding to 142 pc at the absorber.}
\label{model_2cld}
\end{figure}

To account for the absorption spectrum we first adopt a simplified
model, illustrated in Fig.~{\ref{model_2cld}}, 
comprised of one uniform H I `cloud' toward the jet (cloud 1)
and another uniform H I `cloud' toward the core (cloud 2). A single
cloud model is
\noindent  ruled out by past VLBI observations of the 839.4 MHz
absorption feature that indicated an inhomogeneous 
structure of the absorbing gas. The velocity
difference between the maximum depths of the phase-shift
and fringe amplitude spectra detected by
a two-element VLBI experiment (Wolfe {\etal} 1976) 
indicates a velocity difference
between the H I towards the  core and jet sources.
As a result the relative change in Stokes $I$ 
parameter at velocity $v$ is given by

\begin{equation} \label{deltaI}
{\Delta}{I(v)/I}=\sum_{i=1}^{n_{c}}C_{i}f_{i}{\Biggl[}{\rm exp}{\biggl (}-{\tau_{i}(v){\biggr )}}-1{\Biggr]}
\end{equation}

\noindent where the number of clouds $n_{c}$=2, $C_{1}$ and $C_{2}$ are
the area covering
factors of the gas toward the jet and core, $f_{1}$ and $f_{2}$
are the fractions of Stokes $I$ flux density in the jet and
the core sources, and ${\tau_{1}(v)}$ and ${\tau_{2}(v)}$ are the
21 cm optical depths  averaged across the areas subtended
by jet
and core at the absorber. Note, $I$ in the denominator of eq. (4)
is the continuum Stokes $I$ parameter extrapolated across
the 21 cm absorption line.

In order to fit the model to the data we made the following
assumptions (see Fig.~{\ref{model_2cld}}):
First, because most of the polarized flux density arises
from the jet, \footnote{Because the source is weakly polarized,
the flux from the  jet 
is dominated by 
Stokes $I$ with smaller contributions
from Stokes $U$ and $V$ comprising the polarized contribution.}
we assume the absorption spectrum of polarized flux
density, $P({v})$, arises
only in `cloud' 1. Since the line center of 
the $P({v})$ spectrum is shifted by $+$
5.0$\pm$0.25 kHz ($-$1.8$\pm$0.1 km s$^{-1}$)
relative to the line center of the Stokes
$I({v})$ spectrum, we assume that the velocity centroid
of cloud 1 is lower than that of cloud 2. Second,
the
shift in polarization position angle with 
frequency , $\Delta \chi({v})$$\equiv$${\chi}({v})$
$-$ $\chi_{\rm cont}$ (where $\chi_{\rm cont}$ is the polarization position
angle  averaged over the continuum source), indicates
that `cloud' 1 partially covers source 1, i.e., $C_{1}$
$<$ 1, and the intrinsic value of
$\chi$ varies across the source, as is observed in
VLBI experiments at high frequencies 
(Jiang {\etal} 1996; Cotton {\etal} 1997). With a covering
factor $C_{1}$ $<$ 1, the position angle of polarized
flux behind the cloud is weighted less in the line
than in the continuum, thereby causing a shift of position angle
in the line.
Third, because of the smaller dimensions of the core
source, we assume $C_{2}$=1. Fourth, we assume Gaussian
velocity distributions for each cloud, in which case
the optical depth of the  $i^{\rm th}$ cloud at velocity 
$v$ is given by

\begin{equation} \label{tauv}
\tau_{i}(v)={\tau_{0,i}}{\times}{\rm exp}{\Biggl [}-{\biggl (}(v-v_{i})/{\sqrt 2}{\sigma_{v,i}}{\biggr )}^{2}{\Biggr ]}
\end{equation}

\noindent where $v_{i}$  and
and $\sigma_{v,i}$  are the velocity centroid and velocity
dispersion
of the $i^{th}$ cloud.  
In the case of 21 cm absorption the central optical depth

\begin{equation} \label{tau0i}
{\tau}_{0,i}={{{N}_{\rm HI,i}}/ {\biggl (}{4.57{\times}10^{18}{\sigma}_{v,i}}{T_{s,i}}}{\biggr )}
\end{equation}

\noindent where $T_{s,i}$ is the spin temperature and $\sigma_{v,i}$
is in units of km s$^{-1}$. Fifth,
because the polarized radiation is assumed to be emitted only
by the jet source, the relative change in fractional polarization
is given by

\begin{equation} \label{DelP}
{\Delta P(v)}/P=C_{1}{\Biggl [}{\rm exp}{\biggl (}-{\tau_{1}}(v){\biggr )}-1{\Biggr ]}.
\end{equation}

\noindent where we again extrapolate across the line to obtain
$P$ in the denominator of the last equation. The geometry
of the model is illustrated in Fig. \ref{model_2cld}

\subsubsection{Model Fits To The Data}

We adopted an incremental approach by first modeling the
$\Delta I/I$ spectrum alone. When a successful fit was found we
then fitted $\Delta I/I$ together with either the $\Delta P/P$ or $\Delta \chi$
spectrum simultaneously. 
If a successful fit were obtained, we followed up by fitting
$\Delta I/I$, $\Delta P/P$, and $\Delta \chi$ spectra simultaneously. 
The 1-$\sigma$ errors, which were derived by computing
the standard deviations from portions of the spectra displaced from 
the absorption features, are as follows: $\sigma_{\Delta I/I}$=0.00050,
$\sigma_{{\Delta P}/P}$=0.0034, 
and $\sigma_{\Delta \chi}$=0.042$^{\circ}$.

\begin{figure}[b!]  
\begin{center}
\includegraphics[width=3.0in]{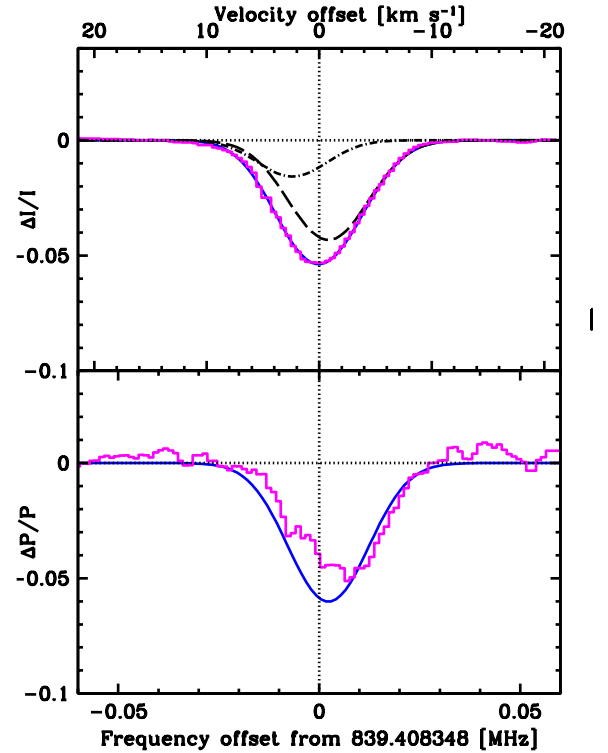} 
\end{center} 
\vspace{-5mm}
\caption{Stokes
spectra
of 3C 286. Solid curves are two-cloud fits and data given by
histograms. Top  panel shows $\Delta I/I$ spectrum. 
Spectrum with dashed curve showing solution for cloud 1 and
dot-dashed curve for cloud 2. 
Bottom panel
shows  $\Delta P/P$.}
\label{Stokes_2cld}
\end{figure}

We obtained the relevant parameters by fitting the model to the data
with the Levenberg-Marquardt method that makes
use of the non-linear chi square minimization
routine, $mrqmin$ (Press {\etal} 1996).
For the  ${\Delta I}/I$ spectrum we fixed the area covering fractions
$C_{1}$=0.5 and $C_{2}$=1.0  as discussed
above, but allowed the other parameters to float after making initial
guesses that were guided by the discussion in $\S$ 5.2.1.
Specifically we solved for $N_{\rm H I,i}/T_{s,i}$
\footnote{Since we make no {\it a priori} assumptions about the
value of 
$N_{\rm H I,i}$ in either cloud, we cannot solve for
$N_{\rm H I,i}$ nor $T_{s,i}$ separately.} 
, $v_{i}$, $\sigma_{\rm v,i}$,
and $f_{i}$. 
A successful fit was found
and the 
output parameters were then used as  inputs 
for fitting
the $\Delta I/I$ and $\Delta P/P$ spectra simultaneously. The
results are shown in Fig.~{\ref{Stokes_2cld}}.

While the fit to the $\Delta I/I$ spectrum appears to be
quite accurate,
the fit including  the $\Delta P/P$ spectrum
has a  chi square per degree of freedom, $\chi^{2}/{\nu}$=12.6, 
which is unacceptable. The reasons
for the failure of the model are straightforward.
Both the depth of the model $\Delta P/P$ absorption feature
and the location of its frequency centroid are in significant
disagreement with the data. Altering the input parameters
to get a better fit to the $\Delta P/P$ spectrum does not
solve the problem, because  such changes 
also alter the $\Delta I/I$ spectrum, which
result in significant
disagreement between the
model and observed $\Delta I/I$ spectra. 
The dilemma is that while cloud 1 produces  both the
$\Delta P/P$ spectrum and most of the $\Delta I/I$
spectrum, the difference between the velocity centroids
of these spectra are too large to be modeled with a
single velocity component.
Because
of a similar difference between
the frequency centroids of the $\Delta \chi$ and $\Delta I/I$
spectra, 
it is clear that the
polarized radiation is incident on   a third velocity component.
As a result
we abandon the two-cloud model
in favor of a three-component model to which we now turn.

%\newpage

\subsection{Three Component Model}.

\begin{figure}[b!]  \begin{center}
\includegraphics[width=3.0in]{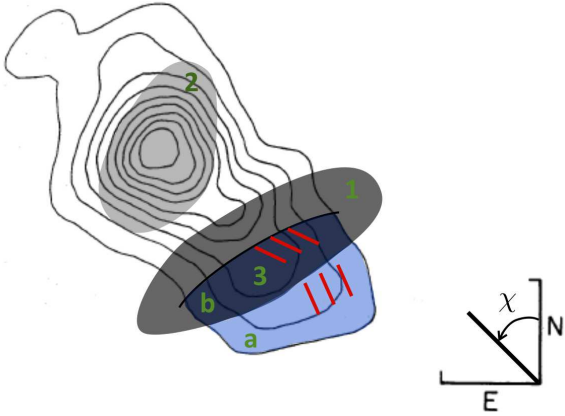} 
\end{center} 
\caption{Same as
{Fig.~{\ref{model_2cld}}} except that cloud 1 
covers both unpolarized (white) and polarized (blue)
portions of the jet. Gas in cloud 1
toward polarized portion is called component 3, which has a  
velocity centroid lower than that of 
the entire cloud 1.  
Note that intrinsic polarization
position angle $\chi$ (where $\chi$ is defined to increase
positively in counter-clockwise
direction from due North) decreases along the source in the SW
direction. $\chi=\chi_{b}$ for the  area
of the polarized jet
behind cloud 1 (i.e., behind  component 3)
and $\chi=\chi_{a}$ for the area of the jet not behind
cloud 1.
Orthogonal straight
lines are 0.02 arcsec in length corresponding to 142 pc at the absorber.} 
\label{model_3cld}
\end{figure}

For this model we again ignore polarized radiation from the core
and assume that it is
emitted only by the jet. 
\footnote{Although the core
is observed to be polarized at higher frequencies
[e.g. Jiang {\etal} 1996], in $\S$5.1  we argued
that the flux density of polarized radiation mainly
comes from the jet.}
Again we assume that
most of the flux from the jet is in Stokes $I$. 
But we now assume that part of the jet 
has low or vanishing polarization and that `cloud' 1 covers
both this unpolarized portion of the jet and that part of the
jet which  emits polarized radiation (the
faint blue section of the source in 
 Fig.~{\ref{model_3cld}}):
the covering factor for
the latter is 
$C_{3}$. We further assume the presence of a velocity gradient
in `cloud' 1 such that the velocity centroid for the Stokes $I$
spectrum is again $v_{1}$, but for the $\Delta P/P$ spectrum is
now $v_{3}$ (see Fig.~{\ref{model_3cld}}).
As a result we now have
\begin{equation} \label{deltap3}
{\Delta P(v)}/P=C_{3}{\Biggl [}{\rm exp}{\biggl (}-{\tau_{3}}(v){\biggr )}-1{\Biggr ]}.
\end{equation}

\noindent where $\tau_{3}(v)$ is the 21 cm optical depth
of the gas in cloud 1 toward the polarized portion of 
the jet source: the velocity centroid of this gas is $v_{3}$.
Therefore, this model has velocity components
centered on $v_{1}, v_{2}, {\rm and} \  v_{3}$.

Since we shall also consider fits to the 
$\Delta \chi$ spectrum in this case, we develop a model 
for the intrinsic change in polarization position angle
across the jet source.
We
assume that $\chi=\chi_{a}$ for the polarized flux density emitted by
the fraction of the jet source that is {\em not} incident on
the cloud and $\chi=\chi_{b}$ for the fraction of emitted polarized
flux density that is incident on 
the cloud. Noting that the net flux densities in 
Stokes $Q$ and $U$ parameters are
given by $Q=Q_{a}+Q_{b}$ and $U=U_{a}+U_{b}$, and
$\chi$=0.5{$\times$}arctan($U/Q$), we find that

%\begin{equation} \label{arctan}
%{\chi}={1 \over 2}{\rm arctan}{\Biggl [}{{C_{1}{\times}r{\times}{\rm tan}(2{\chi_{b}}){\times}{\rm exp}(-\tau_{1}(v))+(1-C_{1}){\times}{\rm tan}(2{\chi_{a}})} \over {C_{1}{\times}r{\times}{\rm exp}(-\tau_{1}(v))+1-C_{1}}}{\Biggr ]}
%\end{equation}

\begin{equation} \label{arctan}
{\chi}(v)={1 \over 2}{\rm arctan}{\Biggl [}{r{C_{3}{\rm tan}(2{\chi_{b}}){\rm exp}(-\tau_{3}(v))+(1-C_{3}){\rm tan}(2{\chi_{a}})} \over {rC_{3}{\rm exp}(-\tau_{3}(v))+1-C_{3}}}{\Biggr].}
\end{equation}

\noindent Here 
$r$=$<{\cal I}_{Q}>_{ b}$/$<{\cal I}_{Q}>_{a}$, where
$<{\cal I}_{Q}>_{ a}$ and $<{\cal I}_{Q}>_{b}$ are the
surface brightnesses in Stokes $Q$ averaged over regions a and b respectively,
and we make use of the relation
$Q_{b}/Q_{a}=rC_{3}/(1-C_{3})$.
The observed
value of $\chi$ averaged over the  unattenuated continuum source,
$\chi_{\rm cont}$, 
is obtained
from the last equation by setting $\tau_{3}$=0. The difference between
$\chi$ in the line and
$\chi_{\rm cont}$ is denoted by $\Delta \chi(v)$$\equiv$$\chi-\chi_{\rm cont}$.

We fitted the three-component model parameters to the spectra as follows.
First, observing that 
the parameters characterizing the Stokes $\Delta I/I$ spectrum
are independent of those characterizing 
the $\Delta P/P$ and $\Delta \chi$ spectra,
we did not fit the three Stokes spectra simultaneously. Rather,
we  fitted the  $\Delta I/I$ spectrum alone, and then the $\Delta P/P$
and $\Delta \chi$ spectra simultaneously. Given the reasonable
fit to the $\Delta I/I$ spectrum discussed in $\S$ 5.2, we adopt
the parameters found for clouds 1 and 2 in that fit here.
Second, in the case of the polarized spectra we need to find 
the gas parameters, $N_{{\rm HI},3}/T_{s,3}$, $v_{3}$,
$\sigma_{v,3}$, and $C_{3}$ as well as the source parameters $r$,
$\chi_{a}$, and $\chi_{b}$. For the source
parameters, we are guided by single-dish polarization measurements
above 1.5 GHz
that show the continuum polarization position angle for
the entire source $\chi_{\rm cont}$=33$\pm$5$^{o}$ (Tabara
and Inoue 1980). The
VLBI polarization data  at 5 GHZ (Jiang {\etal}
1996; Cotton {\etal} 1997; Cotton 2010 priv. comm.)
further show  that  along the jet
$\chi$ decreases with increasing
distance 
from the core source (i.e., in the SW direction) by
as much as $\approx$ 40$^{\circ}$ 
which is consistent with the orientation of the
polarization bars in Fig~{\ref{model_3cld}}.   
We found that the $\Delta \chi(v)$ line spectral feature 
is more naturally explained by a
large central optical depth $\tau_{0,3}$ combined with
a low covering factor $C_{3}$ rather than vice versa.
By contrast, $C_{3}$ and $\tau_{0,3}$ are degenerate in
the case of the $\Delta P/P$ spectrum (see eq. (8)).

\begin{table*}[t!] 
%\tabletypesize{\tiny}
{\scriptsize
\begin{center}
\begin{tabular}{lccccccccc}
\cline{2-10}
&\multicolumn{5}{c}{Gas Properties}&\multicolumn{4}{c}{Source Properties}  \\
\cline{2-5} \cline{7-10}
Cloud  &${\tau_{0,i}}$&{\nh}$_{,i}$/$T_{s,i}$&$v_{i}$$^{a}$&$\sigma_{v,i}$&&$C_{i}$& $f_{i}$& $\chi_{a}$&$\chi_{b}$   \\
    &&cm$^{-2}$K$^{-1}$ &km s$^{-1}$&km s$^{-1}$&     &      &  &   deg. & deg. \\
\tableline
1      & 0.128$\pm$0.005 & (2.03$\pm$0.01)${\times}10^{18}$& $-$0.82$\pm$0.02 & 3.48$\pm$0.02 &  & 0.5  & 0.72   &  ....    &    ....   \\
2      & 0.057$\pm$0.005 & (7.94$\pm$0.10)${\times}10^{17}$& $+$2.37$\pm$0.07 & 3.05$\pm$0.07 &  & 1.0  & 0.28   &  ....    &    ....   \\
3      & 0.280$\pm$0.004 & (4.40$\pm$0.02)${\times}10^{18}$& $-$1.44$\pm$0.04 & 3.42$\pm$0.04 &  & 0.2$^{b}$  & $>$0.40$^{c}$ &  2.70$\pm$0.9    &    43.6$\pm$0.9   \\

\end{tabular}
\end{center}
\caption{Properties of Clouds and Source Causing the Absorption Feature} \label{data}
%\tablenotetext{a}{Velocity $v$=0 corresponds to $z$=0.69215109, shown
%as dotted vertical lines in Figs.1,4, and 6.}
%\tablenotetext{b}{Fraction of polarized flux density incident on ``cloud 3''.}
%\tablenotetext{c}{Fractional area of jet source emitting  polarized radiation, assuming
%$r$=1.}
{\bf (a)} Velocity $v$=0 corresponds to $z$=0.69215109, shown
as dotted vertical lines in Figs.1,4, and 6. \\
{\bf (b)} Fraction of polarized flux density incident on ``cloud 3''. \\
{\bf (c)} Fractional area of jet source emitting  polarized radiation, assuming
$r$=1.
}
\end{table*}

We used $mrqmin$ to determine the 
cloud 1 and cloud 2 parameters by fits to the
$\Delta I/I$ spectrum. As discussed above, the parameters
of the third velocity component
were found
by separately fitting  the  $\Delta P/P$
and $\Delta \chi$ spectra together. This is possible since 
the polarized spectra are independent of the
cloud 1 and 2 parameters, while the Stokes $\Delta I/I$ spectrum
is independent of the component 3 parameters.
After trials with several values of $C_{3}$ 
we fixed $C_{3}$=0.2 at the outset. Larger values of
$ C_{3}$ must be accompanied by smaller values
in $\tau_{0,3}$ in order to maintain the observed
value of $\Delta P/P$ (see eq. 8),  but 
an increase in $C_{3}$ generates
unacceptably large values of $\chi_{b}-\chi_{a}$ in
order to
retain the maximum 
dip of $\Delta \chi$$\approx$ $-$1.5$^{\circ}$(see eq. 9).  
On the other hand smaller values of $C_{3}$ imply larger values of
the ratio $<{\cal I}_{Q}>_{b}/<{\cal I}_{Q}>_{a}$, i.e. 
larger variations of polarized
surface brightness across the source,
than observed at 5 GHz (Jiang {\etal} 1996; Cotton {\etal} 1997).
We considered letting $C_{3}$ be one of the
\noindent floating variables, but decided to let it
remain fixed, because the spatial variation
of $\chi$ across the jet is at best a qualitative rather
than a robust constraint.
For these reasons
we fixed $r$=1 at the outset (where $r$ is the ratio
of average Stokes $Q$ surface brightnesses in region b relative to a), 
but let the remaining
parameters float.

\begin{figure}[b]  
\begin{center}
\includegraphics[width=3.5in]{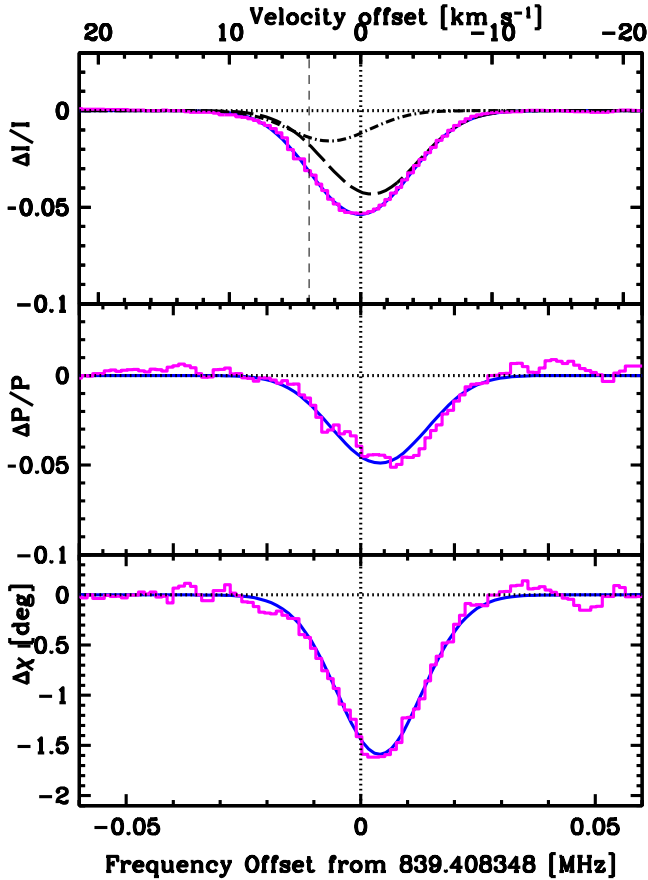} 
\end{center} 
\caption{{\small Stokes
spectra
of 3C 286. Solid blue curves are three-component fits and data are
magenta
histograms. 
Top 
panel shows
$\Delta I/I$
spectrum with cloud 1 and 2 solutions 
as dashed and dot-dashed black curves.
Middle and bottom panels show $\Delta P/P$ and
$\Delta \chi$ spectra. 
Model parameters  are in Table 1.
Vertical dotted line at offsets
=0 corresponds to centroid of Stokes $I$ feature
at $z$=0.69215109$\pm$0.0000000109. Dashed vertical line in top
panel at $v$=4.21$\pm$0.11 km s$^{-1}$ shows
velocity  predicted by the optical redshift 
at $z$=0.69217485$\pm$0.00000058
(Wolfe {\etal} 2008).}} 
\label{Stokes_3cld}
\end{figure}

The results are shown in Fig.~{\ref{Stokes_3cld}},
which plots  model and
observed spectra.
The output parameters are listed in Table 1 along with
1-$\sigma$ errors. These were obtained from
$\Delta {\chi}^{2}$ confidence ellipses containing
68$\%$ of the distributed data by projecting
the ellipses onto each parameter axis (see Press {\etal} 1996).
We did this for all possible pairs of the parameters in
Table 1 and chose the largest value of the  projection
as the 1-$\sigma$ error. 
While the fit to the $\Delta I/I$ spectrum
appears to be quite good, 
the chisquare per degree of freedom  $\chi^{2}/{\nu}$=1.73,
which corresponds to a probability of $\approx$10$^{-8}$ that ${\chi}^{2}$ exceeds
294 for 169 degrees of freedom. One reason for the
elevated ${\chi}^{2}/{\nu}$ is that our model is
undoubtedly inadequate given the $\approx$ 100:1  signal-to-noise
ratio of the $\Delta I/I$ data.
In principle we could improve the fit by adding 
more
velocity components. But 
this would be an unconstrained {\it ad hoc}
procedure, given our limited knowledge of
the source-gas configuration, and thus we
decided against it.
The 
main reason for
the high value of ${\chi}^{2}/{\nu}$ is that the
largest contribution to ${\chi}^{2}$ comes from the
relatively large scatter of the data around the
model between 839.385 MHz  
and 839.397 MHz (i.e., between $\Delta \nu$=$-$ 0.023 
and $-$ 0.011 
MHz in Fig. 6). As 
the cause of this effect is currently uncertain, we attribute
it to some unknown systematic error.

The fits to the polarized spectra result in
$\chi^{2}/{\nu}$=1.54. This  
corresponds to a probability of 
$\approx$ 3{$\times$}10$^{-10}$ that ${\chi}^{2}$ exceeds
536 for 348 degrees of freedom. The reasons for the  high
value of ${\chi}^{2}/{\nu}$ in this case stems from the small,
$\approx$ 3 kHz, difference
between the velocity centroids of the $\Delta P/P$ and $\Delta \chi$
spectra. In contrast to the velocity difference between
the centroids of the polarized and Stokes $I$ spectra, this 
effect is difficult to understand. In any plausible model
the velocity centroids  of the
$\Delta P/P$ and  $\Delta \chi$ spectra coincide with the
centroid of the optical-depth velocity distribution (see eqs. 4, 8
and 9). We tested the possibility that the error
might be due to a constant frequency offset between 
the $\Delta P/P$ and $\Delta \chi$ profiles
by shifting the frequency scale of the $\Delta P/P$
spectrum. We found that $\chi^{2}/{\nu}$ decreased
to a minimum of 1.48 when the shift was one frequency pixel (i.e., 1.2 kHz).
However, the improvement of the fit was insufficient
for the error to be a frequency offset.
More likely this is also an unknown systematic error. 
Adding another velocity component doesn't help since
though it might improve
the fit to the $\Delta P/P$ spectrum at 839.405 MHz,
it also worsens the fit to the $\Delta I/I$ spectrum. 
Despite these problems, the model provides an adequate
fit to the data, given that our goal is to determine
a general picture of the cloud structure rather than
obtain a precise evaluation of cloud sizes, locations,
etc. 
As a result
we shall adopt the three-component model in further discussions.

\section{DISCUSSION} \label{discussion}

In this section we discuss implications of the models
used to describe the 21 cm absorption in 3C 286.

\subsection{Cloud and Galaxy Properties} 

In $\S$ 5.3 we discussed the three-component model used to
model the $\Delta I/I$, $\Delta P/P$, and $\Delta \chi$
absorption spectra. As illustrated in Fig.~{\ref{model_3cld}} 
the three-component model is actually
based on two clouds. One cloud  (cloud 2) completely
covers the core source. The other cloud (cloud 1) partially
covers the jet source, and contains a velocity gradient
such that the part of the cloud toward the polarized portion
of the jet (component 3) causes the velocity centroid of
the polarized spectra to be offset
in velocity from the unpolarized  $\Delta I/I$ spectrum. 
We simplified the model by letting component 3 be a
uniform sub-region of cloud 1 with a velocity centroid, velocity
dispersion, and 21 cm optical depth  that differ from
their values in the parent cloud.
%Support for this model stems from the qualitative agreement
%between the predicted  and observed variation of polarization
%position angle across the jet. 
Table 1 indicates
that the polarization position angle for the part of the source 
absorbed by the foreground cloud is $\chi_{b}$=43.6$^{\circ}$.
Because the ratio of the Stokes parameters, $U/Q$=tan(2$\chi$),
the predicted ratio $U/Q$=20.4. By contrast, $U/Q$=0.046 
toward the part of the jet that bypasses the absorbing gas where
the model predicts $\chi_{a}$=2.67$^{\circ}$. This is 
consistent with the independent Stokes $U$ and $Q$ absorption
spectra, which show a clear detection of the
absorption feature in Stokes $U$ but not in Stokes $Q$.
The absorption feature, which is 
detected at a signal-to-noise ratio of about 20:1 in Stokes $U$,
is predicted to be below the 1-$\sigma$ noise level in Stokes $Q$.
The model also predicts 
$\chi_{cont}$=38$^{\circ}$ for the position angle integrated over the
entire source, which is consistent
with single-dish measurements of  33$\pm$5$^{\circ}$ at 1.5 GHz
(Tabara \& Inoue 1980). 
Our model predicts that the intrinsic 
polarization position angle of the jet
decreases by $\chi_{a}-\chi_{b}$=
$-$40.9$^{o}$ in the SW direction
(see Fig.~{\ref{model_3cld}}). 
Polarization maps at 5 GHZ (Jiang {\etal} 1996; Cotton {\etal} 1997)
do show
shifts by about this amount in the sense of
decreasing $\chi$ along the jet axis in the SW direction. However,
the value of $\chi$ at the extreme edge of the jet appears
to be greater than our prediction of $\chi_{a}$=2.67$^{\circ}$. 
These predictions
could be checked with VLBI polarization measurements
at lower standard frequencies such as 609 MHz.

Next we turn to the properties of the absorbing
gas.
The large projected areas subtended by the sources 
at the absorber
($A_{\rm core}$$\approx$70${\times}$35 pc$^{2}$ for the core 
and $A_{\rm jet}$$\approx$280${\times}$60
pc$^{2}$ for the jet) likely indicate that the 
gas causing 21 cm absorption in 3C 286 is comprised of many interstellar
clouds. In the Galaxy a beam with area $A_{\rm jet}$ 
directed perpendicular to the plane 
of the disk would subtend $\cal N_{\rm CNM}$$\approx$1000 CNM
(i.e., cold neutral-medium [Wolfire {\etal} 1995])
clouds. This is because 
$\cal N_{\rm CNM}$=$n_{CNM}$${\times}A_{\rm jet}$$\times$2$H$ where  the space
density of CNM clouds $n_{CNM}$ $\approx$ 3$\times$10$^{-4}$ pc$^{-3}$ 
(McKee \& Ostriker 1977)
and the disk half-thickness $H$ $\approx$ 100 pc. This value of 
$\cal N_{\rm CNM}$ is a lower limit, since
$\cal N_{\rm CNM}$=$n_{CNM}$$\times$$A{\times}$2$H$/cos($i$) for
a disk with inclination angle $i$.
%{\bf $<<<$Carl this is based on standard CNM cloud model. It would be
%great if could improve on this with your non-spherical cloud
%model$>>>$}
For the core source   
$\cal N_{\rm CNM}$$\ge$60. As a result the Gaussian shape of
the absorption profile is naturally explained 
as the optical-depth weighted sum of multiple
Gaussians (i.e., the central limit theorem). 

Note
that the central velocities of the Gaussians are
likely superposed on  large-scale velocity gradients
present both in `cloud' 1 toward the jet and `cloud' 2
toward the core. While such a gradient in `cloud' 1
is required to explain the velocity shifts between
the stokes $I$ and polarized spectra, as explained
above, the necessity for  a gradient in `cloud' 2 stems
from the approximate agreement between the optical
redshift and redshift of `cloud' 2 illustrated in Fig. 6.
This implies that the optical continuum source is, 
as expected,
physically associated with the compact radio core.
The small, but significant, difference between these
redshifts further suggests that since the optical
beam size is small compared to the dimensions
of `cloud' 2, the optical beam samples only
a limited
portion of the velocities spanned by the gradient
in `cloud' 2. Therefore, although the UV resonance lines
form in gas within `cloud' 2, the velocity centroids
formed by averaging over the optical and wider radio beams
need not be equal.

To determine whether the absorbing gas is CNM, WNM,
or something else
we need to determine its kinetic temperature, $T_{k}$.
Assuming the foreground gas covers the entire source
with a uniform cloud
characterized by $\sigma_{v}$ = 3.75 km s$^{-1}$
(Wolfe {\etal} 2008), one finds an upper
limit of $T_{k}^{\rm max}$=1690 K.
If we further assume that
{\nh} equals the   
UV-determined value of 1.77{$\times$}10$^{21}$ cm$^{-2}$(e.g.
Wolfe \& Davis 1978; Kanekar 2003a), then eq. (6)
implies
$T_{s}$=1035 K, since the central 21 cm optical
depth $\tau_{0}$=0.10 (Wolfe {\etal} 2008). 
Because $T_{s} \ < T_{k} \ < T_{k}^{\rm max}$ for diffuse
warm gas (Liszt 2001), these results would indicate the
gas is neither standard CNM where $T_{k}{\approx}$ 80 K nor
standard WNM where $T_{k}$ $\approx$ 8000 K (Wolfire {\etal} 1995),
but rather resembles the thermally unstable phase 
found by Heiles \& Troland (2003) in the Galaxy ISM. 

On the
other hand the frequency-dependent change in polarization position angle 
across the  absorption feature (panel 3 in Fig.~{\ref{model_3cld}})
provides
strong evidence against the assumption  of
a single uniform cloud (see $\S$ 5.4).
From the values of $\sigma_{v,i}$ in Table 1 
we find that $T_{k,i}^{\rm max}$=1450 K, 1115 K, and 1400 K
for gas in velocity components 1, 2, and 3. If we again assume
that {\nh}$_{,i}$=1.77{$\times$}10$^{21}$ cm$^{-2}$ in
each case, the values of {\nh}$_{,i}$/$T_{s,i}$
in Table 1 indicate that
$T_{s,1}$=892 K 
$T_{s,2}$=2230 K, and
$T_{s,3}$=394 K.
However, since the values of
{\nh}$_{,i}$/$T_{s,i}$ in Table 1 are spatial averages over
dimension much larger than the $\approx$ 1 lt. yr. size of the
UV continuum,  it is unlikely that the UV
determined value of {\nh}  
applies to the gas in each component, nor that
each component has the same value of {\nh}. 
It is equally 
plausible to assume that {\nh} averaged
across cloud 1 is a factor of 2 lower than
the UV determined value, and  $C_{1}$ 
equals 0.25 rather than 0.5. In that case 
$T_{s,1}$=223 K. Similarly if
{\nh}$_{,3}$={\nh}$_{,1}$,  then $T_{s,3}$=197 K. Since these 
are physically plausible temperatures for
CNM gas with the low metal abundances inferred
for this absorber (Wolfe {\etal} 2008), it is reasonable
to assume they represent kinetic temperatures.  Interestingly,
the assumption {\nh}$_{,2}$=1.77{$\times$}10$^{21}$
cm$^{-2}$ results in the unreasonable prediction
that $T_{s,2}$ is higher than
$T_{k,2}^{\rm max}$: we take this as
direct evidence
that {\nh}$_{,2}$ is lower than 1.77{$\times$}10$^{21}$
cm$^{-2}$ and/or $C_{2}$ $<$ 1, thereby implying
that cloud 2 could also be CNM.
Of course, we cannot entirely rule out the possibility that the gas
is in a thermally
unstable phase. But the above arguments
in addition to the results from 
the VLBI line experiment, which  indicates 
individual components toward the core and jet with
kinetic temperatures $T_{k}$ $<$ 500 K (Wolfe {\etal} 1976),
make a compelling case
that the gas  causing 21 cm absorption
in 3C 286
is mainly CNM. Although this conclusion is at odds
with the anti-correlation between $T_{s}$ and [M/H]
found by Kanekar {\etal} (2009b), the evidence
provided by our polarization measurements suggests
either that physical conditions in this absorber
deviate from typical conditions in DLAs or that most
values of $T_{s}$ deduced from UV-determined
values of {\nh} are poor indicators of $T_{k}$.

Identification of the galaxy hosting the absorbing gas
is difficult to ascertain. Le Brun {\etal} (1997) used
HST images to tentatively identify an object (2c in
their nomenclature) as the leading candidate for the
host. Although its absolute magnitude $M_{B}=-20.1$ (in our
cosmology) suggests a massive galaxy, the metallicity
[M/H]=$-$1.6 of the absorbing gas (Wolfe {\etal} 2008) is
at the low end for this redshift,  which suggests
a low-mass galaxy (Tremonti {\etal} 2004). 
Although
the area subtended by 3C 286 at the absorber would
encompass over 1000 CNM clouds, the 3.75 km s$^{-1}$ velocity dispersion
of the  absorption feature more closely resembles
that of a single cloud rather than
an ensemble of clouds. But this might be explained by
a reduced supernovae input of mechanical energy into gas
comprised of many clouds. This 
is consistent with the  low metallicity and absence of
{\ciis} absorption both of which indicate a lower
than normal SFR (Wolfe {\etal} 2008).  
As a result, the evidence accumulated so far suggests
that the gas causing 21 cm absorption
in 3C 286 is embedded in a massive galaxy with a low SFR.

\subsection{Limits on Variations of Physical Constants}

 The difference between the optical redshift, $z_{\rm opt}$, and
radio redshift, $z_{\rm radio}$, places limits on variations of
physical constants between the absorption epoch and the
present, which is given by 

\begin{equation}
{\Delta}{\ln}({\alpha}^{2}g_{p}{m/M})=|(z_{\rm opt}-z_{\rm radio})/(1+z_{\rm radio})|,
\end{equation}

\noindent Here $\alpha$ is the fine structure constant, $g_{p}$
is the gyromagnetic ration of the proton, and $m/M$ is the
electron-to-proton mass ratio (Wolfe {\etal} 1976). Because the radio
redshift
$z_{\rm radio}$=0.69215109$\pm$0.0000000109 and the optical
redshift 
$z_{\rm opt}$=0.69217485$\pm$0.00000058, 
${\Delta}{\ln}({\alpha}^{2}g_{p}{m/M})$=4.2 {\kms} $\pm$ 0.10 {\kms}
(in velocity units) where the
error is dominated by the uncertainty in $z_{\rm opt}$. However, 
this estimate ignores the much larger   systematic error
determined from night-to-night changes in wavelength calibration
on HIRES, which can be as large as 2 {\kms} (Kanekar {\etal} 2010). 
Therefore,  adopting the 4.2
{\kms} difference as a conservative 2$\sigma$  upper limit, we find 
${\Delta}{\ln}({\alpha}^{2}g_{p}{m/M})$ $<$ 1.4$\times$10$^{-5}$ in 
the redshift interval $z$=[0, 0.692].
On the other hand our conclusion
that the optical source is located in the compact
core source implies that the relevant radio redshift is 
given by cloud 2, which is shifted
by 1.8 {\kms} from $z_{\rm opt}$ (see Table 1). In that case
${\Delta}{\ln}({\alpha^{2}}g_{p}{m/M})$ $<$ 0.6$\times$10$^{-5}$.
Both of these limits are comparable to results from recent measurements by
Kanekar {\etal} (2010; see also Tzanavaris {\etal} 2007) who used redshifts deduced
from   C I absorption  to determine the optical redshift, since
C I is a more accurate tracer of the CNM gas that gives rise
to 21 cm absorption than the Si II, Fe II, Zn II, and Cr II resonance transitions
used to determine the optical redshift of the 21 cm
absorber toward 3C 286 (Wolfe {\etal} 2008). In principle
the present limit  could be improved with a future space based detection  
of C I absorption lines in this absorber,
and a reduction of the systematic errors in
wavelength calibration of the HIRES spectrograph.

\section{SUMMARY AND CONCLUSIONS} \label{summary}

Reanalysis of the polarization spectra of the 21 cm
absorption line detected at $z$=0.692 toward 3C 286
leads to the following conclusions:

(1) The detection of an 84 {\muG}
magnetic field inferred from Zeeman splitting  by
Wolfe {\etal} (2008) is erroneous.
An insidious software error failed to remove a spurious phase 
delay of $\approx$ 90$^{\circ}$ between orthogonal linearly
polarized signals from the GBT antenna. This phase
difference changed the derived Stokes $V$ parameter into
Stokes $U$ and vice versa. As a result, what we interpreted
as the spectral signature of Zeeman splitting in Stokes $V$
was a shift between the velocity centroids of the
Stokes $I$ and Stokes $U$ absorption spectra.

(2) A proper reanalysis of the GBT spectra is presented
in Fig. 1. 
The Stokes $I$ spectrum again exhibits the
21 cm absorption feature at $\nu$ =
839.408348 MHz ($z$=0.692115109). 
The spectrum of fractional polarization, 
p$\equiv$${\sqrt{Q^{2}+U^{2}}}/I$,
where $Q$ and $U$ are the Stokes parameters, shows
that the frequency centroid of 
the absorption feature is
offset by  5.0$\pm$0.25 kHz from the frequency centroid
of the Stokes $I$ spectrum. 
The polarization position-angle spectrum exhibits a frequency
centroid that is also offset from the frequency
centroid of the Stokes $I$ spectrum, but by an
amount that differs slightly from the
centroid of the fractional polarization
spectrum.
The Stokes $V$ spectrum exhibits an absorption feature
at 839.41 MHz, but this is most likely an artifact
due to leakage from the other Stokes parameters. The
latter imply a 3-$\sigma$ upper limit of 17 $\mu$G on
the line-of-sight component of the magnetic field
in the absorbing gas.

(3) We modeled the  absorption
feature with a core-jet radio structure
of 3C 286 behind  three velocity components: 
in $\S$ 5.2.2 we describe why two velocity
components will not work.
In the three-component  model, cloud 2 completely covers the core source
and 
cloud 1 partially covers the jet with a sub-region
3 that partially covers the polarized portion of the
jet. 
The radio beam through cloud 1 has a covering
factor $C_{1}$ for the Stokes $I$ radiation,
but a smaller covering factor $C_{3}$ for the polarized
radiation (see Fig.~{\ref{model_3cld}}). Since
we assume the presence of a velocity gradient in this
cloud the velocity centroids in Stokes $I$
and the polarized radiation in the jet will differ.
%Furthermore, the model predict that polarization position angle
%decreases with increasing displacement from the core
%source (Fig.~{\ref{model_3cld}} ). 
The predictions of this model
are in reasonable agreement with the data.
Comparison between the model and observed spectra
is shown in Fig.~{\ref{Stokes_3cld}, and the physical parameters
of the model are given in Table 1.

(4) In \S 7 we discuss implications of these models.
We show that the change in polarization position
angle predicted by the model is consistent with
polarization measurements made at higher frequencies,
but with some important differences that may
be tested with future VLBI observations. We
show that the area of the jet projected onto
the absorbing gas 
would encompass more than 1000 standard
CNM type clouds found in the Galaxy. This provides
a natural explanation for the Gaussian shape of the
absorption feature through the central limit
theorem. We argue that the high spin temperatures
deduced from the single uniform cloud model are
unlikely to be correct due to evidence for
an inhomogeneous distribution of the foreground gas.
While we cannot rule out 
the possibility that we have detected warm, thermally
unstable gas,
we have presented independent arguments favoring the CNM
hypothesis.
The evidence accumulated so far also indicates the
absorbing gas is in a massive galaxy with a lower
than average SFR.

(5) A comparison between the radio and optical redshifts
of the 21 cm absorber sets a conservative 2$\sigma$
upper limit  on variations on the product of three physical
constants given by
${\Delta}{\ln}({\alpha}^{2}g_{p}{m/M})$ $<$ 1.4$\times$10$^{-5}$
within the redshift interval $z$=[0,0.692].
Based on our conclusion that  the optical absorption occurs in
cloud 2,  this limit reduces to 
${\Delta}{\ln}({\alpha}^{2}g_{p}{m/M})$ $<$ 0.6$\times$10$^{-5}$.

To conclude, our measurement of 21 cm absorption
in all four Stokes parameters has provided new
information about the foreground gas at $z$=0.692 toward
3C 286. The  difference between the velocity
centroids of the absorption feature in
unpolarized radiation on the one hand and
in fractional polarization and polarization
position angle on the other hand demonstrates
evidence for  spatial variations of {\nh}/$T_{s}$
on scales less than $\sim$ 100 pc. Our observations
demonstrate that simple models of uniform clouds
covering the entire radio source structure are
incorrect. This is clearly illustrated by our
analysis of cloud 2 in front of the compact core
source: application of the UV-determined value
of {\nh} leads to a value of the spin temperature 
significantly higher than the upper limit to the
kinetic temperature set by the velocity dispersion
of the gas, which is physically implausible.
Instead the data are more consistent with a model
in which each of the three velocity components
is comprised of large numbers of small CNM
clouds that produce the smooth Gaussian velocity
profile of the absorption feature.

{\bf Acknowledgements}:
We wish to thank Kim Griest for valuable discussions
concerning data analysis and Marc  Rafelski
for constructing Figs.~{\ref{vlbimap}}, {\ref{model_2cld}},
and {\ref{model_3cld}}. The GBT is one of the
facilities of the 
National Radio Astronomy
Observatory, which is a center of the National
Science Foundation under cooperative agreement by
Associated Observatories Inc..
AMW and JXP are partially supported by NSF grant AST07-09235.
CH and TR are  partially supported by NSF grant AST09-08841.

%\end{thebibliography}{}
%\bibitem[Wolfe\ (1995)]{wolfe95}
%Wolfe, A. M., Lanzetta, K. M., Foltz, C. B., \& Chaffee, F. H.  1995,
%\apj, 454, 698
%\end{thebibliography}


\begin{references}

\reference {} Boisse, P., Le Brun,V. Bergeron, J. \& Deharveng, J.-M,
A{\&}A, 333, 841


\reference {} Cotton, W. D., Fanti, C., Fanti, R., Dallacasa, D.,
Foley, A. R., Schilizzi, R. T., \& Spencer, R. E., 1997, A{\&}A, 325, 479

\reference {} Heiles, C.  2001, PASP, 113, 1243

\reference {} Heiles, C., Perillat, P., Nolan, M, Lorimer,
  D., Bhat, R., Ghosh, T., Lewis, M.,  O'Neil, K.,
  Salter, C., \& Stanimirovic, S. 2001, PASP, 113, 1274

\reference {} Heiles, C., Troland, T. H. 2003 ApJ, 586, 1067

\reference {} Heiles, C., Troland, T. H. 2004 ApJS, 151, 271

\reference {} Jiang, D. R., Dallacasa, D., Schilizzi, R., T., Ludke, E.,
Sanghera, H. S., \& Cotton, W. D. 1996, A{\&}A, 312, 380

\reference {} Kanekar, N., Lane, W. M., Momjian, E., Briggs, F. H.,
\& Chengalur, J. N. 2009a, MNRAS, 394, L61

\reference {} Kanekar, N., Smette, A., Briggs, F. H.,
\& Chengalur, J. N. 2009b, ApJ, 705, L40

\reference {} Kanekar, N., Prochaska, J. X., Ellison, S. L.,
\& Chengalur, J. N. 2010, ApJL, 712, L148

\reference {} Kronberg, P. P., ApJ, 676, 70

\reference {} Le Brun, V., Bergeron, J., \& Deharveng, J. M. 1997, AA,
321, 733

\reference {} Liszt, H. 2001, A{\&}A, 370, 698

\reference {} McKee, C. F., \& Ostriker, J. P 1977, ApJ,  218, 148

\reference {} Press, W. H., Teukolsky, S., A., Vetterling, W. T.,
\& Flannery, B. 1996, Numerical Recipes in FORTRAN 77, (Cambridge:
University Press), p. 680

\reference {} Simon, R. S., Readhead, A. C. S., \& Moffet, A. 1980, ApJ, 236, 
707.

\reference {} Tabara, H., \& Inoue, M. 1980, A{\&}A Sup., 39, 379 

\reference {} Tremonti, C. A;., Heckman, T. M., Kauffmann, G., Brinchmann, J.,
Charlot, S., White, S. D. M., Seibert, M., Peng, E. W., Schlegel, D. J.,
Uomoto, A., Fukugita, M., \& Brinkmann, J. 2004, ApJ, 613, 898

\reference {} Tzanavaris, P., Webb, J. K., Murphy M. T., Flambaum, V. V.,
\& Curran, S. J. 2007, MNRAS, 374, 634

\reference {} Wilkinson, P. N., Readhead, A. C. S., Anderson, B., 
{\&} Purcell, G. H. 1997, ApJ, 232, 365


\reference {} Wolfe, A. M., Broderick, J. J., Condon, J. J., \& 
Johnston, K. J. 1976. ApJ, 208, L47

\reference {} Wolfe, A. M., Brown, \& R. L., Robert, M. S. 1976,
PhRvL., 37, 179.

\reference {} Wolfe, A. M., \& Davis, M. M. 1978, AJ, 84, 699

\reference {} Wolfe, A. M., Gawiser, E., \& Prochaska, J. X. 2005,
ARAA, 43, 861

\reference {} Wolfe, A. M., Jorgenson, R. A., Robishaw, T., Heiles, C.,
\& Prochaska, J. X. 2008, Nature, 455, 638. 

\reference {} Wolfire, M. G., McKee, C. F., Hollenbach, D., \&
Tielens, A.G.G.M. 1995, ApJ, 453, 673
\end{references}
\end{document}